\documentclass[baaa]{baaa}

\usepackage[pdftex]{hyperref}
\hypersetup{colorlinks=true,
			linkcolor=blue,
			filecolor=blue,
			urlcolor=blue,
			citecolor=blue,
			}
\usepackage{subfigure}
\usepackage{natbib}
\usepackage{helvet,soul}
\usepackage[font=small]{caption}

\begin{document}


\journalvol{61A}
\journalyear{2019}
\journaleditors{R. Gamen, N. Padilla, C. Parisi, F. Iglesias \& M. Sgr\'o}


\contriblanguage{1}


\contribtype{2}

\thematicarea{4}

\title{Dynamo effect in the double periodic variable}
\subtitle{DQ Velorum}


\titlerunning{Dynamo effect in the DPV DQ Vel}


\author{R. I. San Mart\'in-P\'erez\inst{1}, D. R. G. Schleicher\inst{1}, R. E.  Mennickent\inst{1}, \& Rosales. J.~A.\inst{1}}
\authorrunning{San Mart\'in-P\'erez et al.}


\contact{rubensanmartinp@gmail.com}

\institute{Departamento de Astronom\'ia, Facultad de Ciencias F\'isicas y Matem\'aticas, Universidad de Concepci\'on, Av. Esteban Iturra s/n, Barrio Universitario, Casilla 160-C, Chile}


\resumen{DQ Velorum es una variable de doble per\'iodo gal\'actica la cual consiste de un sistema binario semi-separado con una gainer de tipo espectral B y una estrella donor de tipo espectral A, adem\'as de un disco de acreci\'on en la estrella gainer. El sistema presenta un per\'iodo orbital de $6.08337$~d\'ias y un per\'iodo largo de $189$~d\'ias cuyo origen sigue en debate. En este trabajo estudiamos la posibilidad de que este per\'iodo sea producto de un d\'inamo magn\'etico estudiando la evoluci\'on completa del sistema. El modelo ajusta de muy buena manera el estado actual del sistema y puede ser usado para describir la evolucion de DQ Velorum. Adem\'as, el modelo predice un incremento en el n\'umero de d\'inamo en la estrella donor en \'epocas de alta transferencia de masa y una raz\'on de los per\'iodos largo y orbital muy cercana a la que se observa en el sistema.}

\abstract{DQ Velorum is a galactic double periodic variable (DPV), this system is a semi-detached binary comprised of a B-type gainer and an A-type donor star plus an extended accretion disc around the gainer. The system also presents an orbital period of $6.08337$~days and a long period of $189$~days whose origin is still under debate. Here we studied the possibility that this period may be driven by a magnetic dynamo investigating the entire evolution of the system. The model matches in a very good way the current state of the system and it can potentially be used to describe the evolution of DQ Velorum. It also  predicts an increase of the dynamo number of the donor during epochs of high mass transfer in this system, and a theoretical long/orbital period ratio very close to the observed one at the present system age.}



\maketitle

\section{Introduction}
\label{S_intro} 
An important class of close interacting binaries are the so-called Algol-type variables. Such systems consist of semi-detached binaries with intermediate mass components. In these systems the less massive star (donor) is more evolved than the most massive star (gainer) and the mass ratio of these systems indicates that some processes may have occured in order to explain the high mass of the main sequence companion star, as in particular mass transfer as originally explained by \citet{crawford1955} and confirmed by \citet{kippenhahn, vanrensbergen2011} and \citet{demink2014} through numerical calculations. The apparent mass paradox between binary components can be well understood if the donor star was the most massive star in the system, it evolved first and started a fast process of mass exchange through Roche lobe overflows (RLOF) onto its companion \citep{eggleton2006}.

\indent One of the most important features in the Algol-type binaries is the presence of long cycles. The presence of such cycles is known since a long time (e.g. \citet{lorenzia1980, lorenzib1980, guinan1989}). Nevertheless, the interpretation and origin of these long cycles are still under debate.\\

\indent We aim to give an explanation on the origin of this cycle. We focused our study on a sub-class of the Algol classification which are the double periodic variable (DPV) binary systems which consist of semi-detached interacting binary systems with intermediate mass components that were found by \citet{mennickent2003} presenting the main characteristics of the Algol systems.

This report is based on the DPV system DQ Velorum, a binary system that was fully studied by \citet{barriab2013, barria2013, barria2014} which presents the following stellar parameters: M$_{d}=$ $2.2\pm 0.2$~M$_{\odot}$, M$_{g}=$ $7.3\pm 0.3$~M$_{\odot}$, R$_{d}=$ $8.4\pm 0.2$~R$_{\odot}$, R$_{g}=$ $3.6\pm 0.2$~R$_{\odot}$, L$_{d}=$ $2.66\pm0.036$~L$_{\odot}$, L$_{g}=$ $3.14\pm0.16 $~L$_{\odot}$, P$_{orb}=$ $6.08337\pm 0.00013$~days and a long period of $P_{long}=189$~\,days. Here we explored the magnetic dynamo cycle as a potential origin of the long period. For this purpose we studied the entire evolution of this systems using the binary evolution models proposed by \citet{vanrensbergen2008, vanrensbergen2011} and for the dynamo we used the relation proposed by \citet{soon1993} and \citet{baliunas1996} that relates the rotational velocity, activity period and the dynamo number $D=\alpha\Delta\Omega d^{3}/\eta^{2}$. Here $\alpha$ is a measure of helicity, $\Delta \Omega$ is the large-scale differential rotation, $d$ is the characteristic length scale of convection and $\eta$ the turbulent magnetic diffusivity in the star.

\section{Methods}
In order to study how the dynamo number $D$ and the ratio of the long to orbital period changes as the system is evolving we are fitting our system to the van Rensbergen binary evolution models \citep{vanrensbergen2008, vanrensbergen2011}. These models were derived in order to study the evolution of close binaries including both conservative and non-conservative scenarios by solving the stellar structure equations using the \textit{Brussels} binary evolutionary code (for a detailed description of the code see \citet{deloore1992}). The code was modified in order to include convective mixing, radius corrections and nuclear physics, following \citet{prantzos1986}. Also, following \citet{delooredegreve1992} a moderate convective core overshooting is applied. Mass loss by stellar winds and period changes due to angular momentum loss are also included in this model. Initial conditions were established by using an unevolved system with a B-type primary at birth from the 9th catalog of spectroscopic binaries \citep{pourbaix2004}, distinguishing between late B-type $[2.5, 7]$~M$_\odot$ and early B-type $[7,16.7]$~M$_\odot$ primaries. We inspected all the 561 conservative and non-conservative evolutionary tracks that are available at the Center Done\'es Stellaires (CDS) looking for the model that describes the best our system. A multiparametric $\chi^{2}$ minimization was performed in order to find the best match, this test is given by \citep{mennickent2012}:
\begin{eqnarray}
\chi^{2}_{i,j}=\left(\frac{1}{N}\right)\sum_{k}w_{k}\left[\frac{(S_{i,j,k}-O_{k})}{O_{k}}\right]^{2},
\label{eq: chi}
\end{eqnarray}
where $N$ is the number of observations $(7)$, $S_{i,j,k}$ is the synthetic model where $i$ indicates the model, $j$ the time $t_{j}$ and $k$ the stellar or orbital parameter, $O_{k}$ are the observed stellar parameters. To perform our test, we are fitting the mass, radii and luminosities of both stars in the system and also the orbital period of the binary. $w_{k}=\sqrt{O_{k}/\epsilon O_{k}}$ is the statistical weight of the parameter $O_{k}$ and $\epsilon O_{k}$ is the error associated to the parameter $O_{k}$. The model with the minimum $\chi^{2}$ corresponds to the model that describes best the evolution of the system. After we found the best model for our system, we are following the relation between the long period $P_{long}$  and the orbital period $P_{orb}$ given by \citep{soon1993, baliunas1996}
\begin{eqnarray}
P_{long}=D^{\gamma}P_{rot},
\end{eqnarray}
with $D$ the dynamo number that measures how magnetically active a star is and $\gamma$ a power law index with values, usually between $1/3$ and $5/6$ \citep{saar1999, dube2013}. In order to calculate the dynamo number $D$ we follow the dynamo model proposed by \citet{schleicher2017} where they proposed that the long period is related to the orbital period following the relation

\begin{eqnarray}
P_{long}=&P_{orb}\left(11.5\left(\frac{2\sqrt{2}}{15}\right)^{1/3}\frac{\mathrm{R_{\odot}}}{\mathrm{km\, s^{-1}\, yr}}\right)^{-2\gamma} \nonumber \\
& \times\left(\frac{L_{2}^{2/3}R_{2}^{2/3}}{M_{2}^{2/3}}\left(\frac{l_{m}}{H_{p}}\right)^{-4/3}\left(\frac{P_{kep}}{\epsilon_{H}R_{2}}\right)^{2}\right)^{-\gamma},
\label{eq: dn}
\end{eqnarray}

where $l_{m}$ is the mixing length, $H_{p}$ is the pressure scale height and $P_{kep}$ is the Keplerian orbital period of a test particle on the surface of the donor star. They found a good fit with $\gamma \sim 0.31\pm 0.05$. More recently \citet{navarrete2018} and \citet{voelschow2018} showed that a magnetic dynamo can explain the modulation periods which are produced by changes in the quadrupole moment of the star in post common envelope (PCE) systems. \\

\subsection{Model fitting}
\indent We applied the $\chi^{2}$ minimization test to the parameters listed in table (\ref{tb: parameters}) and we found that the model that describes best the system corresponds to the same non-conservative and slow mass transfer rate as the one reported by \citet{barria2014} and it presents the following ratios between observations and model predictions: $M_{d, obs}/M_{d, model}=0.94$, $M_{g, obs}/M_{g, model}=1.007$, $R_{d, obs}/R_{d, model}=0.97$, $R_{d, obs}/R_{d, model}=0.99$, $L_{d, obs}/L_{d, model}=1.11$, $L_{g, obs}/L_{g, model}= 0.93$ and $P_{orb, obs}/P_{orb, model}=0.99$ at a stellar age of $\sim 68\pm 0.5$~Myr. This means that the model can describe in a very good way the evolution of the system, the top panel in figure (\ref{fig: fit}) shows the evolutionary track of both components of the system, the path with circles shows the evolution of the gainer star. We also indicate the observational and predicted state of the gainer (black dot with/without error bars, respectively). The path with triangles shows the evolution of the donor star, also it is plotted the observational state of the donor (red triangle with error bars) and the prediction of the model for the donor (red triangle with no error bars). The prediction of the current state of each component of the system almost overlaps with the current observational state of the binary.

\subsection{Evolution of the dynamo number and long-to-orbital period ratio}
\begin{figure}[!h]
  \centering
  \includegraphics[width=0.35\textwidth]{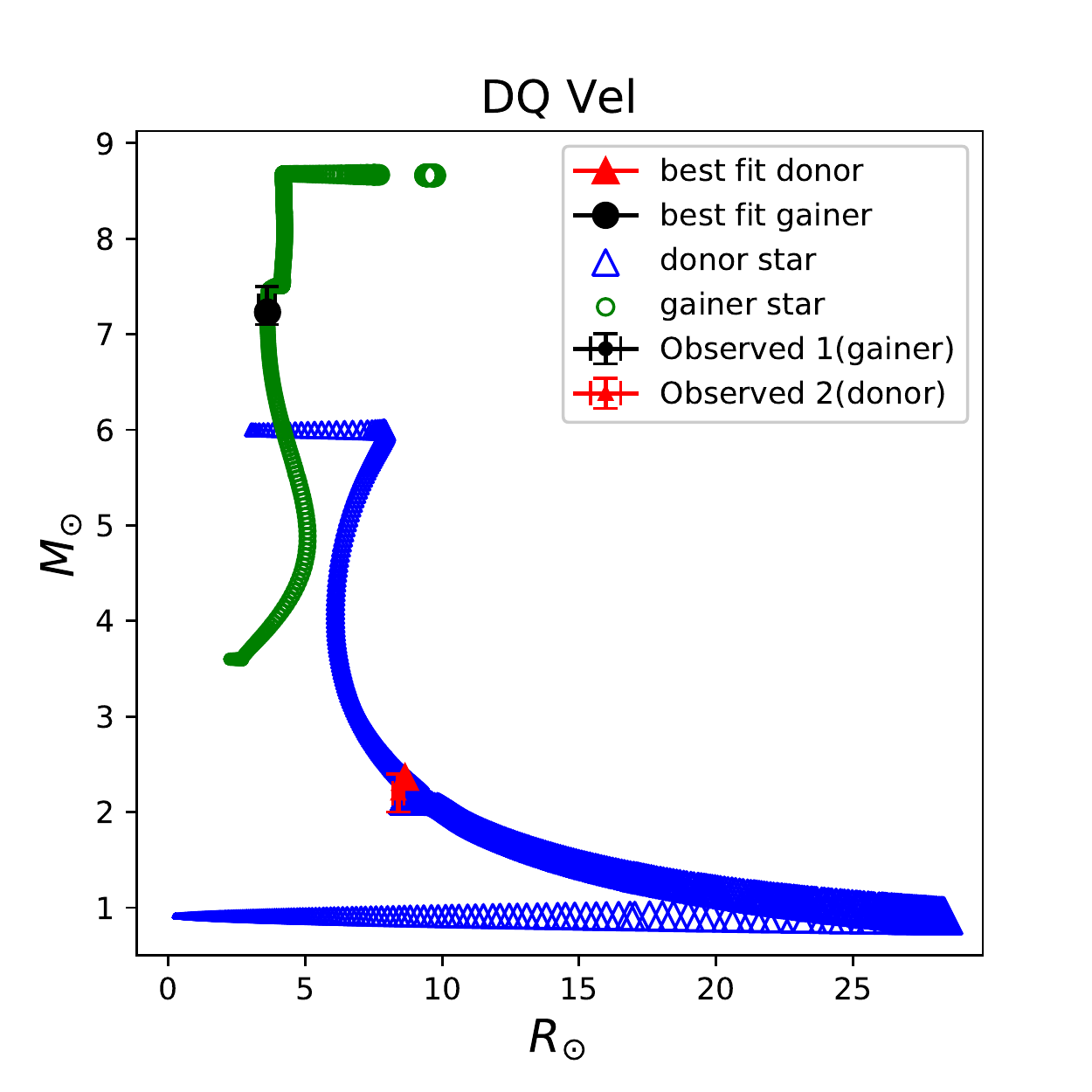}
  \includegraphics[width=0.4\textwidth]{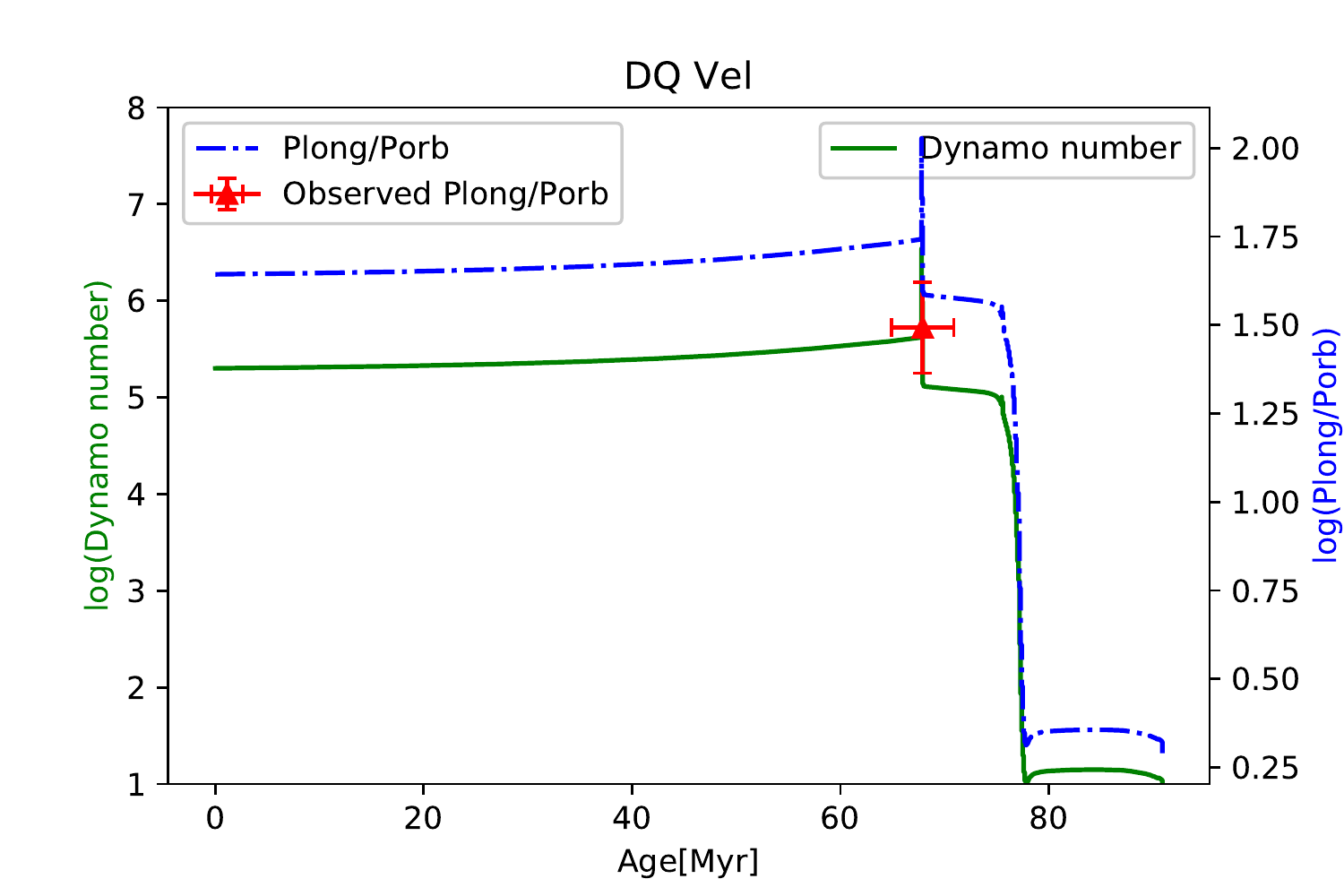}
  \includegraphics[width=0.4\textwidth]{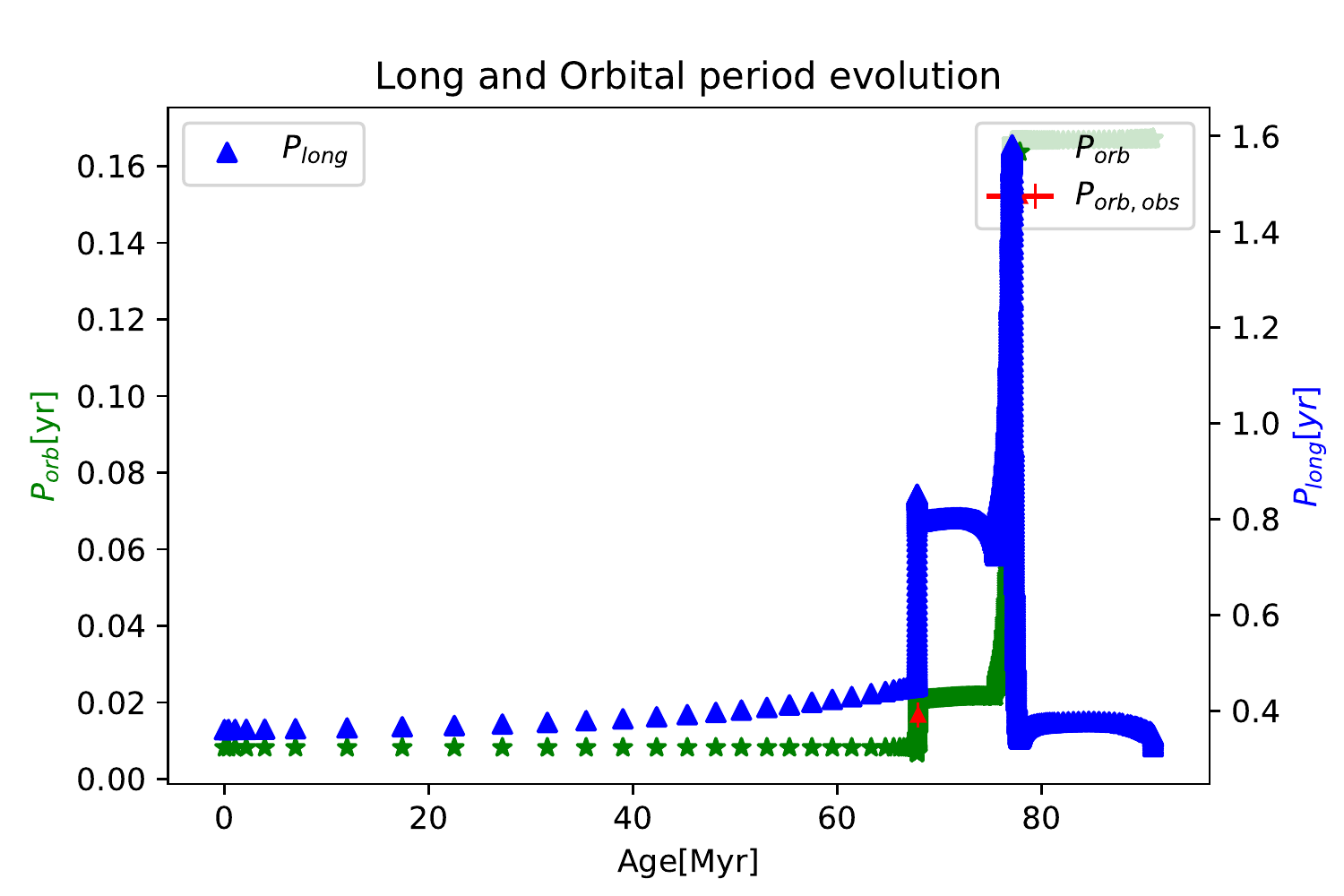}
  \caption{(Top): The best model for DQ Vel. We show
the evolution of the gainer (circles), its observed  (black dots with errors) and predicted (red triangles) parameters. The path with triangles shows the evolution of the donor, the observed (red triangle with error bars) and predicted
parameters (red triangle without error bars). (Middle): The dynamo number (solid line) and the ratio of long to orbital period (dashed line). We also plot the observed ratio of long to orbital period (triangle with error bars) with a value of $\log(P_{long}/P_{orb}) = 1.492$, the same value reported by \citet{schleicher2017}. (Bottom): The evolution of the orbital period (path with asterisks) and the long period (path with triangles) and the observed value of the orbital period marked with an error-bar triangle in log scale.}
  \label{fig: fit}
\end{figure}

Once the fit is done both dynamo number $D$ and ratio of long to orbital period were calculated using equation (\ref{eq: dn}). As seen in figure (\ref{fig: fit}) middle, the dynamo number $D$ follows an almost constant evolution during the first $55$~\,Myr, after this the dynamo number grows dramatically fast, which will possibly produce a strong magnetic activity and the stars begins an active mass transfer phase. This situation lasts for a very short time ($\sim 1- \sim 2$)~kyr, and right after the dynamo peak, it starts to decrease until the star becomes magnetically inactive at a stellar age of $\sim 77$~Myr. The ratio of long to orbital period also grows as fast as the dynamo number, this means that both periods (long and orbital) are almost constant during the first $55$~\,Myr, after this constant evolution, both periods grows very rapidly producing a star with a longer magnetic cycle and with a longer orbital period as seen in figure (\ref{fig: fit}) bottom panel. 
\begin{table}[!h]
\centering
\caption{Observational stellar parameters for  DQ Velorum. Taken from \citet{barria2013, mennick2016}}
\begin{tabular}{lcc}
\hline\hline\noalign{\smallskip}
\!\!Parameter & \!\!\!\!Value\!\!\!\!\\
\hline\noalign{\smallskip}
\!\!M$_{d}$~\,[M$_{\odot}$]  & $2.2\pm 0.2$\\
\!\!M$_{g}$~\,[M$_{\odot}$]& $7.3\pm 0.3$\\
\!\!R$_{d}$~\,[R$_{\odot}$]& $8.4\pm 0.2$\\
\!\!R$_{g}$~\,[R$_{\odot}$] &$3.6\pm 0.2$ \\
\!\!L$_{d}$~\,[L$_{\odot}$] &$2.66\pm0.036$ \\
\!\!L$_{g}$~\,[L$_{\odot}$]& $3.14\pm0.16$\\
\!\!P$_{orb}$~\,[days]& $6.08337\pm 0.00013$\\

\hline
\label{tb: parameters}
\end{tabular}
\end{table}

\section{Conclusions}
We have found that DQ Velorum is an old system with a small mass transfer rate as reported by \citet{barria2014}. The system is well described by the models proposed by \citet{vanrensbergen2008, vanrensbergen2011}, so it seems feasible to use it as a potential description for the evolution of the system. We studied the evolution of the dynamo number $D$ finding that after an almost constant evolution of this parameter the star becomes strongly active in a short period due to the short convective time scale in the stellar interior, starting a rapid mass transfer phase, also other stellar parameters such like radius, mass and luminosity are changing fast. This, however, lasts for a very short time ($\sim 1- \sim 2$)~kyr, and after this strongly active phase of the star, the magnetic activity decreases until the star becomes magnetically inactive with no mass exchange between both stars.

\begin{acknowledgement}
DRGS and RISMP thank for funding through Fondecyt regular (project code 1161247) and the ``Concurso Proyectos Internacionales de Investigaci\'on, Convocatoria 2015'' (project code PII20150171). DRGS and RM thank for funding through BASAL Centro de Astrof\'isica y Tecnolog\'ias Afines (CATA) PFB-06/2007. 
\end{acknowledgement}


\bibliographystyle{baaa}
\small
\bibliography{biblio}
 
\end{document}